\documentclass[a4paper,aps,prd,10pt,preprintnumbers,showpacs,showkeys,twocolumn,superscriptaddress,nofootinbib]{revtex4-1}
\usepackage{orcidlink,graphicx,cmap,amsmath}
\usepackage[utf8]{inputenc}
\usepackage[T1]{fontenc}
\usepackage{booktabs}
\usepackage{hyperref}

\begin{document}

\title{Ringing Regularity: Gravitational Perturbations and Quasinormal Modes of Einasto-Supported Black Holes}

\author{Bekir Can Lütfüoğlu}
\email{bekir.lutfuoglu@uhk.cz}
\affiliation{Department of Physics, Faculty of Science, University of Hradec Králové, Rokitanského 62/26, 500 03 Hradec Králové, Czech Republic}

\author{Javlon~Rayimbaev} 
\email{javlon@astrin.uz}
\affiliation{Institute of Theoretical Physics, National University of Uzbekistan, Tashkent 100174, Uzbekistan}
\affiliation{University of Tashkent for Applied Sciences, Gavhar Str. 1, Tashkent 700127, Uzbekistan}


\author{Sardor~Murodov} 
\email{mursardor@ifar.uz}
\affiliation{New Uzbekistan University, Movarounnahr Str. 1, Tashkent 100000, Uzbekistan}
\affiliation{Institute of Fundamental and Applied Research, National Research University TIIAME, Kori Niyoziy 39, Tashkent 100000, Uzbekistan}

\author{Mardon~Abdullaev} 
\email{mardonabdullaev@gmail.com}
\affiliation{Kimyo International University in Tashkent, Shota Rustaveli street 156, Tashkent 100121, Uzbekistan}
\affiliation{Tashkent State Technical University, Tashkent 100095, Uzbekistan}

\author{Munisbek Akhmedov} \email{munisbek95@urdu.uz} \affiliation{Urgench State University, Kh. Alimjan Str. 14, Urgench 221100, Uzbekistan}

\begin{abstract}
We investigate axial gravitational perturbations and quasinormal modes of regular black holes supported by an Einasto distribution of matter. Halo matter not only removes the central singularity but also modifies the quasinormal spectrum. We show that the resulting quasinormal spectrum deviates systematically from the Schwarzschild case, with shifts in both oscillation frequencies and damping rates that grow with the halo scale parameter and the Einasto index. In near-extremal configurations, the damping rate can be significantly suppressed, leading to long-lived modes. The effects of regularity and environmental factors on the spectrum are found to be substantially larger than the estimated numerical uncertainty, as confirmed independently by high-order WKB calculations with Padé resummation and time-domain integration. 
\end{abstract}


\maketitle

\section{Introduction}

The appearance of spacetime singularities in classical general relativity signals a breakdown of the theory in regimes of extreme curvature. According to the singularity theorems, once certain energy and causality conditions are satisfied, gravitational collapse generically leads to geodesic incompleteness. Although such results are robust within classical gravity, they strongly suggest that new physics must intervene at short distances to prevent curvature invariants from diverging. Understanding how singularities are avoided — and whether realistic alternatives to classical black holes can arise — remains one of the fundamental problems of gravitational theory.

A broad spectrum of regular black-hole models has been developed over the past decades  \cite{Bardeen:1968,Hayward:2005gi,Simpson:2018tsi,Ansoldi:2008jw,AyonBeato:1998ub,Dymnikova:1992ux,Bronnikov:2000vy,Bronnikov:2024izh,Bronnikov:2005gm,Kazakov:1993ha,Modesto:2008jz,Bonanno:2000ep,Lan:2023cvz,Bonanno:2023rzk,Bonanno:2025dry,Spina:2025wxb,Zhang:2024ney,Solodukhin:2025opw,Bueno:2024dgm,Bueno:2025tli,Bueno:2024eig,Frolov:2026rcm,Konoplya:2020ibi,Bronnikov:2006fu,Bronnikov:2003gx,Casadio:2001jg,Konoplya:2024hfg}. Some constructions rely on phenomenological modifications of the metric function near the center, designed so that curvature scalars remain finite while preserving asymptotic flatness \cite{Bardeen:1968,Hayward:2005gi,Simpson:2018tsi}. Other approaches interpret the regular core as being supported by effective matter sources, frequently modeled as nonlinear electrodynamics or anisotropic fluids \cite{Ansoldi:2008jw,AyonBeato:1998ub,Dymnikova:1992ux,Bronnikov:2000vy,Bronnikov:2024izh,Bronnikov:2005gm}. In addition, quantum-gravity-motivated scenarios — including loop quantum gravity, asymptotically safe gravity, and higher-curvature or nonlocal extensions of general relativity — naturally produce geometries in which classical singularities are replaced by regular cores \cite{Kazakov:1993ha,Modesto:2008jz,Spina:2025wxb,Bonanno:2000ep,Lan:2023cvz,Bonanno:2023rzk,Bonanno:2025dry,Spina:2025wxb,Zhang:2024ney,Solodukhin:2025opw,Bueno:2024dgm,Bueno:2025tli,Bueno:2024eig,Frolov:2026rcm}. While these models differ in their underlying motivation, they share the common feature that the central region of the black hole is modified in a way consistent with finite curvature.

An important aspect of regular solutions concerns their astrophysical viability. Black holes observed in nature are not isolated objects but are embedded in complex environments, most notably in the dark-matter halos of galaxies. Observations across multiple scales — from galactic rotation curves to large-scale structure formation — provide compelling evidence for extended dark-matter distributions. Although the microscopic properties of dark matter remain unknown, its large-scale gravitational effects are well described by empirical density profiles such as the Navarro–Frenk–White, Hernquist, and Einasto models \cite{Navarro:1996gj,Bertone:2005xz,Hernquist:1990be,Einasto:1965czb}. These phenomenological profiles successfully reproduce galactic dynamics and cosmological simulations without specifying the microphysical equation of state, thereby offering flexible effective descriptions of matter distributions.

It is therefore natural to ask whether a consistent black hole solution can be obtained in the presence of a realistic galactic halo, a problem that has been addressed in a number of recent works~\cite{Cardoso:2021wlq,Konoplya:2025nqv,Konoplya:2022hbl,Dekel:2017bwy,Zhao:1995cp,Kar:2025phe,Sajadi:2023ybm,Sajadi:2025prp}.
Recently, it has been demonstrated that certain dark-matter density profiles can be incorporated directly into exact black-hole solutions in such a way that the matter distribution itself ensures regularity of the geometry. In particular, a family of regular black-hole metrics supported by an Einasto-type distribution was constructed in \cite{Konoplya:2025ect}. In this framework, the same matter profile that describes realistic halo structures also removes the central singularity, providing a natural link between astrophysical environments and regular black-hole geometries.

The dynamical response of such configurations to perturbations is of central importance \cite{Fernando:2012yw,Flachi:2012nv,Li:2013fka,Liu:2020ola,Jusufi:2020odz,Toshmatov:2015wga,Toshmatov:2018ell,Toshmatov:2019gxg,Rincon:2020cos,Yang:2021cvh,Franzin:2022iai,Konoplya:2022hll,Meng:2022oxg,Franzin:2023slm,Konoplya:2023ahd,Bolokhov:2023ruj,Lutfuoglu:2026fks,Skvortsova:2024eqi,Pedrotti:2024znu,Stashko:2024wuq,Skvortsova:2024wly,Khoo:2025qjc,Konoplya:2025uta,Bolokhov:2025egl,Bolokhov:2025lnt,Arbelaez:2025gwj,Malik:2025czt}. Quasinormal modes (QNMs) govern the ringdown phase of perturbed black holes and encode information about the geometry in the strong-field region. They determine the stability of the spacetime and provide potential observational signatures in gravitational-wave signals. For regular black holes, the quasinormal spectrum may deviate from that of the Schwarzschild solution, reflecting modifications to the effective potential near the photon sphere and in the near-horizon region. Therefore, analyzing gravitational perturbations in these geometries is essential for assessing their physical properties and possible observational distinguishability.

In this paper, we investigate gravitational perturbations of the regular black-hole solutions supported by the Einasto density profile found in \cite{Konoplya:2025ect}. We compute the quasinormal spectrum for various values of the profile parameters and analyze how the fitting constants controlling the halo distribution influence the oscillation frequencies and damping rates. This allows us to determine the extent to which astrophysically motivated matter distributions modify the ringdown characteristics of regular black holes and to clarify the interplay between environmental effects and strong-field dynamics.

The paper is organized as follows. In Sec.~\ref{sec:EinastoBackground} we review the construction of spherically symmetric black-hole solutions supported by an Einasto matter distribution and discuss their main geometric properties, including the conditions for the existence of an event horizon and regularity at the center. In Sec.~\ref{sec:perturbations} we derive the equations governing axial gravitational perturbations and present the corresponding effective potentials for the two perturbation sectors. Section~IV is devoted to the computation of QNMs using the high-order WKB method with Pad\'e resummation, supplemented by time-domain integration as an independent check of accuracy. In Sec .~\ref{sec:QNMs}, we also analyze the obtained spectra and discuss their dependence on the Einasto index and the halo scale parameter. Finally, Sec.~\ref{sec:conc} summarizes our results and outlines possible directions for further investigation.

\section{Regular black holes supported by the Einasto profile}\label{sec:EinastoBackground}

We consider static, spherically symmetric geometries described by
\begin{equation}
ds^2 = - f(r)\, dt^2 + \frac{dr^2}{f(r)} + r^2 d\Omega^2 ,
\end{equation}
where the metric function is written in terms of the mass function,
\begin{equation}
f(r)=1-\frac{2 m(r)}{r}.
\end{equation}
The matter source is modeled as an anisotropic fluid with density $\rho(r)$ and radial pressure $P_r(r)$. Following the construction of~\cite{Konoplya:2025ect}, we adopt the effective equation of state
\begin{equation}
P_r(r)=-\rho(r),
\end{equation}
which ensures regularity at the horizon and leads to $B(r)=1$ in the general parametrization of the line element. The mass function then satisfies
\begin{equation}
m(r)=4\pi \int_0^r x^2 \rho(x)\, dx ,
\end{equation}
with $m(0)=0$. Regularity at the origin is guaranteed provided that $\rho(r)$ remains finite and nonnegative for all $r$.

\subsection*{Einasto density profile}

The Einasto profile has become a standard parametric description of galactic and cluster dark-matter halos because it captures a key feature seen in high-resolution $N$-body simulations: the logarithmic density slope is not constant but varies smoothly with radius, i.e., the profile is intrinsically ``curved'' in $\log\rho$--$\log r$ space. This behavior produces systematic (and measurable) improvements over two-parameter cuspy templates such as NFW when fitting simulated halo density and circular-velocity profiles, with the additional shape parameter controlling how rapidly the slope changes with $r$~\cite{Merritt:2006AJ132,Navarro:2003ew}. In the Aquarius simulations of Milky-Way-sized halos, for example, neither host halos nor subhalos exhibit evidence for convergence to a universal inner power-law cusp; instead, their inner structure is well fit by gently curving Einasto forms~\cite{Springel:2008cc}. Moreover, the Einasto shape parameter is not merely a fitting nuisance: it correlates with halo assembly, in the sense that departures from NFW and the preferred Einasto curvature track variations in mass-accretion histories~\cite{Ludlow:2011cs}. Analytic properties of the Einasto model (e.g., enclosed mass, potential, and lensing expressions) have also been worked out in detail, facilitating direct comparison with dynamical and observational constraints~\cite{Retana-Montenegro:2012dbd}.

As a physically motivated example, we consider the Einasto distribution,
\begin{equation}
\rho(r)=\rho_0 \exp\!\left[-\left(\frac{r}{h}\right)^{1/n}\right], 
\qquad n>0 ,
\end{equation}
where $h$ is a scale radius and $n$ controls the curvature of the density slope. This profile is widely used in galactic dynamics and numerical simulations of structure formation because it accurately reproduces the smooth radial distribution of dark-matter halos. The total mass,
\begin{equation}
M=\lim_{r\to\infty} m(r),
\end{equation}
is finite for all $n>0$, so that the geometry is asymptotically Schwarzschild. In the limit $h\to0$ (at fixed $M$), the matter distribution collapses toward the origin and the metric reduces to the Schwarzschild solution.

For generic values of $n$, the integral for $m(r)$ does not admit a simple closed form and must be evaluated numerically. Nevertheless, several important cases are analytic and illustrate the family's main features.

\subsection*{Case $n=1/2$}

For $n=\tfrac12$ the density is Gaussian,
\begin{equation}
\rho(r)=\rho_0 \exp\!\left(-\frac{r^2}{h^2}\right),
\end{equation}
and the mass function can be expressed through the error function,
\begin{equation}
f(r)=1-\frac{2M}{r}\,\mathrm{erf}\!\left(\frac{r}{h}\right)
+\frac{4M}{\sqrt{\pi}\,h} e^{-r^2/h^2},
\end{equation}
with
\begin{equation}
M=\pi^{3/2}\rho_0 h^3 .
\end{equation}

Near the center the metric behaves as
\begin{equation}
f(r)=1-\frac{4M}{3\sqrt{\pi} h^3}\, r^2 + \mathcal{O}(r^4),
\end{equation}
indicating a de Sitter–type core. All curvature invariants remain finite at $r=0$, confirming the absence of a central singularity. For sufficiently small values of $h/M$, the solution possesses two horizons (an event and an inner horizon). As $h$ increases, the horizons merge at an extremal configuration and eventually disappear.

\subsection*{Case $n=1$}

For $n=1$ the density decreases exponentially,
\begin{equation}
\rho(r)=\rho_0 e^{-r/h},
\end{equation}
and the metric function takes the form
\begin{equation}
f(r)=1-\frac{2M}{r}
+M \frac{2h^2+2hr+r^2}{h^2 r} e^{-r/h},
\end{equation}
with
\begin{equation}
M=8\pi \rho_0 h^3 .
\end{equation}

Expanding near the origin gives
\begin{equation}
f(r)=1-\frac{2M}{3 h^3} r^2 + \mathcal{O}(r^3),
\end{equation}
again corresponding to a regular de Sitter–like core. As in the previous case, horizons exist only for sufficiently compact configurations (small $h/M$). The allowed range of $h$ is more restrictive than for $n=\tfrac12$, reflecting the slower decay of the density profile.

\subsection*{General case}

For arbitrary $n$ the qualitative structure is similar. The geometry is asymptotically Schwarzschild at large $r$, while the central region is regular and resembles a de Sitter core determined by $\rho_0$. The existence of horizons depends on the relative compactness of the matter distribution. Black-hole solutions occur only when the scale parameter $h$ is sufficiently small compared to the total mass $M$; otherwise the spacetime describes a horizonless compact object.

The parameter $n$ controls how rapidly the density falls off with radius. Smaller values of $n$ correspond to profiles that decay more rapidly and therefore produce geometries closer to Schwarzschild outside the horizon. Larger $n$ yields a more extended distribution, leading to stronger deviations from the vacuum solution in the strong-field region and a narrower range of parameters admitting an event horizon.

Thus, the Einasto-supported family provides a continuous interpolation between the Schwarzschild geometry and regular configurations with extended matter distributions. In the following sections, we analyze the dynamical properties of these solutions by studying their gravitational perturbations and quasinormal spectra.

\section{Gravitational perturbations and boundary conditions}\label{sec:perturbations}

\begin{figure*}
\resizebox{\linewidth}{!}{\includegraphics{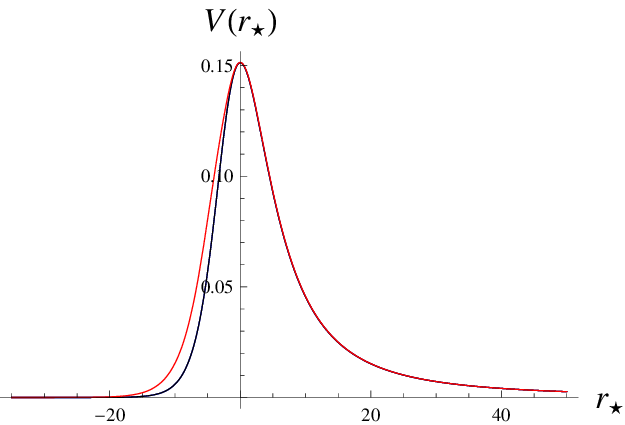}\includegraphics{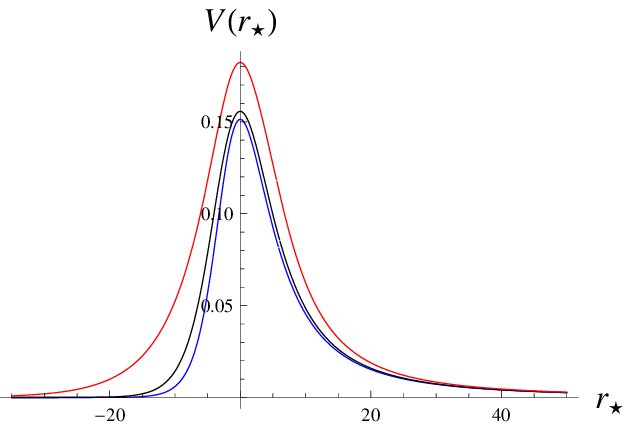}}
\caption{Effective potential as a function of the tortoise coordinate $r^{*}$ for $\ell=2$ up perturbations; $M=1$. Left: $n=1/2$, $h=0.1$ (blue), $h=0.5$ (black) and $h=0.9$ (red). Right: $n=1$, $h=0.1$ (blue), $h=0.3$ (black) and $h=0.38$ (red).}\label{fig:pot}
\end{figure*}

To analyze the dynamical response of the Einasto-supported regular black holes, we consider linear axial (odd-parity) gravitational perturbations. For spherically symmetric backgrounds sourced by an anisotropic fluid, the perturbation equations can be reduced to a single master equation of Schrödinger type. After separating the time dependence as $\Psi(t,r)=e^{-i\omega t}\Psi(r)$ and introducing the tortoise coordinate
\begin{equation}
\frac{dr_*}{dr}=\frac{1}{f(r)},
\end{equation}
the radial part satisfies
\begin{equation}
\frac{d^2 \Psi}{dr_*^2}
+\left(\omega^2 - V(r)\right)\Psi = 0 .
\label{mastereq}
\end{equation}

For anisotropic fluids, there are two inequivalent definitions of axial perturbations depending on how the fluid variables are perturbed. These lead to two distinct effective potentials, which we denote as the ``up'' and ``down'' sectors. In both cases the centrifugal term and the standard $-6m(r)/r^3$ contribution characteristic of gravitational perturbations are present, while additional density-dependent terms reflect the coupling to matter.

The effective potential for the ``up'' sector is \cite{Chakraborty:2024gcr,Konoplya:2025ect},
\begin{eqnarray}
V^{\text{(up)}}(r)
&=&
f(r)\Biggl[
\frac{\ell(\ell+1)}{r^2}
-\frac{6m(r)}{r^3}
+8\pi \rho(r)
\nonumber\\&&
-8\pi r\,\rho'(r)
\Biggr],\label{Vup}
\end{eqnarray}
whereas for the ``down'' sector it reads
\begin{equation}
V^{\text{(down)}}(r)
=
f(r)\left[
\frac{\ell(\ell+1)}{r^2}
-\frac{6m(r)}{r^3}
+8\pi \rho(r)
\right].
\label{Vdown}
\end{equation}

Both potentials reduce to the Regge--Wheeler potential in the vacuum limit $\rho\to0$. For physically reasonable density profiles satisfying $\rho(r)>0$ and $\rho'(r)<0$, the matter contributions are nonnegative outside the horizon. Since $f(r)>0$ for $r>r_0$ and $2m(r)\le r$, the potentials are positive for multipoles $\ell\ge2$. Consequently, the associated differential operator
$
\mathcal{D}
=
-\frac{d^2}{dr_*^2}+V(r)
$ is positive definite, implying the absence of exponentially growing axial modes. Therefore, the solutions are linearly stable in the odd-parity sector \cite{Konoplya:2025ect}.

The effective potentials~(\ref{Vup}) and~(\ref{Vdown}) are displayed in Fig.~\ref{fig:pot}. For small values of $n$, the deviation from the Schwarzschild geometry is confined primarily to the near-horizon region, and the metric rapidly approaches the vacuum solution at larger radii. As $n$ increases, the departure from the Schwarzschild potential extends to progressively larger distances, significantly modifying the region around the peak of the effective potential, which plays a dominant role in determining the properties of quasinormal radiation.

\begin{figure*}
\resizebox{\linewidth}{!}{\includegraphics{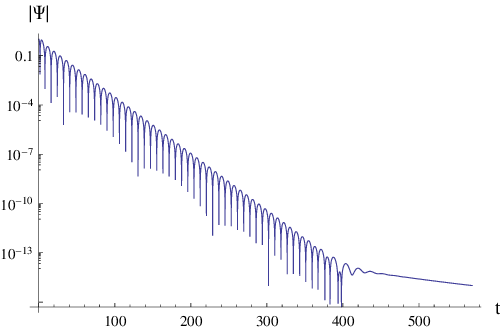}\includegraphics{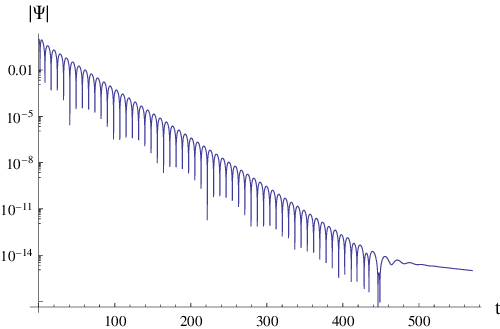}}
\caption{Time domain profile for up-potential $\ell=2$, $n=1$, $h=0.3$. Left: The up-potential for which the 16th WKB result $\omega=0.384390-0.079386 i$ perfectly agrees with the extraction of the frequency from the time-domain profile via the Prony method $\omega = 0.384389 - 0.079382 i$. Right: Down potential for which the 16th WKB result $\omega= 0.382115-0.072757 i$ perfectly agrees with the extraction of the frequency from the time-domain profile via the Prony method $\omega = 0.382115 - 0.0727567 i$.}\label{fig:TD1}
\end{figure*}

\begin{table*}
\centering
\begin{tabular}{l c c l c c l}
\toprule
& \multicolumn{3}{c}{$\ell=2$, $n=1/2$ (up)} 
& \multicolumn{3}{c}{$\ell=3$, $n=1/2$ (up)} \\
\cmidrule(lr){2-4} \cmidrule(lr){5-7}
$h$ 
& WKB-16 ($m=8$) 
& WKB-14 ($m=7$) 
& diff (\%) 
& WKB-16 ($m=8$) 
& WKB-14 ($m=7$) 
& diff (\%) \\
\midrule

0.1  & $0.373676-0.088976 i$ & $0.373671-0.088970 i$ & 0.0022 
     & $0.599443-0.092703 i$ & $0.599443-0.092703 i$ & 0 \\

0.2  & $0.373676-0.088976 i$ & $0.373671-0.088970 i$ & 0.0022 
     & $0.599443-0.092703 i$ & $0.599443-0.092703 i$ & 0 \\

0.3  & $0.373676-0.088976 i$ & $0.373671-0.088970 i$ & 0.0022    
     & $0.599443-0.092703 i$ & $0.599443-0.092703 i$ & 0 \\

0.4  & $0.373671-0.088971 i$ & $0.373660-0.088967 i$ & 0.00283
     & $0.599443-0.092703 i$ & $0.599443-0.092703 i$ & 0 \\

0.5  & $0.373657-0.088960 i$ & $0.373657-0.088960 i$ & 0
     & $0.599443-0.092703 i$ & $0.599443-0.092704 i$ & 0.00007 \\

0.6  & $0.373608-0.088916 i$ & $0.373605-0.088923 i$ & 0.00175
     & $0.599424-0.092700 i$ & $0.599423-0.092703 i$ & 0.00054 \\

0.7  & $0.373231-0.088262 i$ & $0.373232-0.088263 i$ & 0.00042
     & $0.599259-0.092535 i$ & $0.599257-0.092536 i$ & 0.00042 \\

0.8  & $0.371578-0.085598 i$ & $0.371950-0.085174 i$ & 0.148
     & $0.598982-0.091376 i$ & $0.598982-0.091377 i$ & 0.00004 \\

0.9  & $0.374892-0.077986 i$ & $0.374706-0.078472 i$ & 0.136
     & $0.599722-0.087937 i$ & $0.599722-0.087938 i$ & 0.00012 \\

0.95 & $0.365034-0.048147 i$ & $0.362556-0.048347 i$ & 0.675
     & $0.601164-0.085092 i$ & $0.601172-0.085088 i$ & 0.00140 \\

1.0  & $0.354598-0.040244 i$ & $0.354405-0.042711 i$ & 0.693
     & $0.603645-0.081238 i$ & $0.603575-0.081382 i$ & 0.0263 \\

1.05 & $0.382064-0.003834 i$ & $0.381953-0.003900 i$ & 0.0336
     & $0.607674-0.076438 i$ & $0.607677-0.076438 i$ & 0.00038 \\

\bottomrule
\end{tabular}
\caption{Fundamental QNMs for $\ell=2$ and $\ell=3$ ($n=1/2$, up-potential, $M=1$), calculated using the WKB method at different orders with Pad\'e approximants. The last columns show the relative difference between WKB-16 and WKB-14 results (in percent).}
\end{table*}

\begin{table*}
\centering
\begin{tabular}{l c c l c c l}
\toprule
& \multicolumn{3}{c}{$\ell=2$, $n=1/2$ (down)} 
& \multicolumn{3}{c}{$\ell=3$, $n=1/2$ (down)} \\
\cmidrule(lr){2-4} \cmidrule(lr){5-7}
$h$ 
& WKB-16 ($m=8$) 
& WKB-14 ($m=7$) 
& diff (\%) 
& WKB-16 ($m=8$) 
& WKB-14 ($m=7$) 
& diff (\%) \\
\midrule

0.1  & $0.373676-0.088976 i$ & $0.373671-0.088970 i$ & 0.0022
     & $0.599443-0.092703 i$ & $0.599443-0.092703 i$ & 0 \\

0.2  & $0.373676-0.088976 i$ & $0.373671-0.088970 i$ & 0.0022
     & $0.599443-0.092703 i$ & $0.599443-0.092703 i$ & 0 \\

0.3  & $0.373676-0.088976 i$ & $0.373671-0.088970 i$ & 0.0022
     & $0.599443-0.092703 i$ & $0.599443-0.092703 i$ & 0 \\

0.4  & $0.373672-0.088971 i$ & $0.373675-0.088970 i$ & 0.0007
     & $0.599443-0.092703 i$ & $0.599443-0.092703 i$ & 0 \\

0.5  & $0.373668-0.088969 i$ & $0.373667-0.088968 i$ & 0.0005
     & $0.599443-0.092703 i$ & $0.599443-0.092703 i$ & 0 \\

0.6  & $0.373659-0.088969 i$ & $0.373659-0.088969 i$ & 0
     & $0.599441-0.092705 i$ & $0.599441-0.092705 i$ & 0 \\

0.7  & $0.373568-0.088926 i$ & $0.373568-0.088927 i$ & 0.00032
     & $0.599390-0.092710 i$ & $0.599389-0.092711 i$ & 0.00024 \\

0.8  & $0.373230-0.088561 i$ & $0.373196-0.088554 i$ & 0.00887
     & $0.599102-0.092552 i$ & $0.599097-0.092553 i$ & 0.00074 \\

0.9  & $0.372521-0.087170 i$ & $0.372438-0.087146 i$ & 0.0225
     & $0.598432-0.091609 i$ & $0.598431-0.091613 i$ & 0.00069 \\

0.95 & $0.371999-0.085812 i$ & $0.372053-0.085838 i$ & 0.0157
     & $0.598000-0.090543 i$ & $0.598001-0.090543 i$ & 0.00009 \\

1.0  & $0.371346-0.083814 i$ & $0.371197-0.083800 i$ & 0.0394
     & $0.597546-0.088868 i$ & $0.597535-0.088868 i$ & 0.00180 \\

1.05 & $0.370515-0.081113 i$ & $0.371026-0.080705 i$ & 0.172
     & $0.597084-0.086403 i$ & $0.597081-0.086406 i$ & 0.00065 \\

\bottomrule
\end{tabular}
\caption{Fundamental QNMs for $\ell=2$ and $\ell=3$ ($n=1/2$, down-potential, $M=1$), calculated using the WKB method at different orders with Pad\'e approximants. The last columns show the relative difference between WKB-16 and WKB-14 results (in percent). }
\end{table*}

\begin{table*}
\centering
\begin{tabular}{l c c l c c l}
\toprule
& \multicolumn{3}{c}{first overtone, $\ell=3$, $n=1/2$ (down)} 
& \multicolumn{3}{c}{second overtone, $\ell=3$, $n=1/2$ (down)} \\
\cmidrule(lr){2-4} \cmidrule(lr){5-7}
$h$ 
& WKB-16 ($m=8$) 
& WKB-14 ($m=7$) 
& diff (\%) 
& WKB-16 ($m=8$) 
& WKB-14 ($m=7$) 
& diff (\%) \\
\midrule

0.1  & $0.582644-0.281298 i$ & $0.582644-0.281298 i$ & 0
     & $0.551683-0.479094 i$ & $0.551685-0.479091 i$ & 0.00057 \\

0.2  & $0.582644-0.281298 i$ & $0.582644-0.281298 i$ & 0
     & $0.551683-0.479094 i$ & $0.551685-0.479091 i$ & 0.00057 \\

0.3  & $0.582644-0.281298 i$ & $0.582644-0.281298 i$ & 0
     & $0.551683-0.479094 i$ & $0.551685-0.479091 i$ & 0.00057 \\

0.4  & $0.582644-0.281298 i$ & $0.582644-0.281299 i$ & 0.00014
     & $0.551685-0.479089 i$ & $0.551688-0.479099 i$ & 0.00150 \\

0.5  & $0.582642-0.281301 i$ & $0.582644-0.281301 i$ & 0.00022
     & $0.551676-0.479098 i$ & $0.551716-0.479114 i$ & 0.00588 \\

0.6  & $0.582615-0.281307 i$ & $0.582616-0.281317 i$ & 0.00159
     & $0.551707-0.479145 i$ & $0.551470-0.479195 i$ & 0.0332 \\

0.7  & $0.582334-0.281308 i$ & $0.582420-0.281305 i$ & 0.0134
     & $0.550700-0.479150 i$ & $0.551131-0.478915 i$ & 0.0672 \\

0.8  & $0.580921-0.280493 i$ & $0.580926-0.280495 i$ & 0.00071
     & $0.546857-0.476754 i$ & $0.546988-0.476789 i$ & 0.0186 \\

0.9  & $0.578064-0.276701 i$ & $0.577986-0.276886 i$ & 0.0314
     & $0.539666-0.467575 i$ & $0.539546-0.467341 i$ & 0.0367 \\

0.95 & $0.575915-0.272903 i$ & $0.575916-0.272903 i$ & 0.00015
     & $0.534313-0.459660 i$ & $0.534282-0.459467 i$ & 0.0278 \\

1.0  & $0.573080-0.267261 i$ & $0.573041-0.267276 i$ & 0.00665
     & $0.526103-0.448668 i$ & $0.526007-0.449160 i$ & 0.0725 \\

1.05 & $0.569194-0.259748 i$ & $0.569092-0.259839 i$ & 0.0218
     & $0.514674-0.435634 i$ & $0.514241-0.438787 i$ & 0.472 \\

\bottomrule
\end{tabular}
\caption{First and second overtones of QNMs for $\ell=3$ ($n=1/2$, down-potential, $M=1$), calculated using the WKB method at different orders with Pad\'e approximants. The last columns show the relative difference between WKB-16 and WKB-14 results (in percent).}
\end{table*}

\begin{table*}
\centering
\begin{tabular}{l c c l c c l}
\toprule
& \multicolumn{3}{c}{fundamental mode, $\ell=2$, $n=1$ (up)} 
& \multicolumn{3}{c}{first overtone, $\ell=2$, $n=1$ (up)} \\
\cmidrule(lr){2-4} \cmidrule(lr){5-7}
$h$ 
& WKB-16 ($m=8$) 
& WKB-14 ($m=7$) 
& diff (\%) 
& WKB-16 ($m=8$) 
& WKB-14 ($m=7$) 
& diff (\%) \\
\midrule

0.10 & $0.373672-0.088962 i$ & $0.373671-0.088962 i$ & 0.00019
     & $0.346746-0.273898 i$ & $0.346740-0.273944 i$ & 0.0104 \\

0.15 & $0.373659-0.088876 i$ & $0.373657-0.088882 i$ & 0.00163
     & $0.346815-0.273570 i$ & $0.346802-0.273557 i$ & 0.00428 \\

0.20 & $0.373977-0.087854 i$ & $0.373978-0.087854 i$ & 0.00012
     & $0.348064-0.270497 i$ & $0.347981-0.271076 i$ & 0.133 \\

0.25 & $0.376560-0.084655 i$ & $0.376560-0.084656 i$ & 0.00009
     & $0.355109-0.261852 i$ & $0.355123-0.262178 i$ & 0.0740 \\

0.30 & $0.384390-0.079386 i$  & $0.384375-0.079370 i$ & 0.0057   
     & $0.370453-0.247558 i$ & $0.370387-0.247464 i$ & 0.0258 \\ 

0.35 & $0.400786-0.072448 i$ & $0.400582-0.072461 i$ & 0.0500
     & $0.392029-0.225034 i$ & $0.388267-0.220879 i$ & 1.24 \\

0.36 & $0.405679-0.071079 i$ & $0.405621-0.071083 i$ & 0.0139
     & $0.396549-0.223462 i$ & $0.394486-0.224976 i$ & 0.562 \\

0.37 & $0.411223-0.068496 i$ & $0.410653-0.068902 i$ & 0.168
     & $0.396562-0.200005 i$ & $0.382275-0.202594 i$ & 3.27 \\

0.38 & $0.417168-0.066723 i$ & $0.417176-0.066752 i$ & 0.00692
     & $0.390121-0.194392 i$ & $0.390039-0.194425 i$ & 0.0203 \\

\bottomrule
\end{tabular}
\caption{Quasinormal modes of the $\ell=2$, $n=1$ up-potential ($M=1$), calculated using the WKB method at different orders with Pad\'e approximants. The two blocks correspond to the two datasets shown separately above. The last columns display the relative difference between WKB-16 and WKB-14 results (in percent).}
\end{table*}

\begin{table*}
\centering
\begin{tabular}{l c c l c c l}
\toprule
& \multicolumn{3}{c}{fundamental mode, $\ell=3$, $n=1$ (up)} 
& \multicolumn{3}{c}{first overtone, $\ell=3$, $n=1$ (up)}\\
\cmidrule(lr){2-4} \cmidrule(lr){5-7}
$h$ 
& WKB-16 ($m=8$) 
& WKB-14 ($m=7$) 
& diff (\%) 
& WKB-16 ($m=8$) 
& WKB-14 ($m=7$) 
& diff (\%) \\
\midrule

0.10 & $0.599443-0.092703 i$ & $0.599443-0.092703 i$ & 0
     & $0.582643-0.281298 i$ & $0.582643-0.281298 i$ & 0.00002 \\

0.15 & $0.599433-0.092669 i$ & $0.599433-0.092669 i$ & 0
     & $0.582609-0.281161 i$ & $0.582611-0.281163 i$ & 0.00043 \\

0.20 & $0.599586-0.092146 i$ & $0.599586-0.092146 i$ & 0
     & $0.582932-0.279477 i$ & $0.582932-0.279478 i$ & 0.00020 \\

0.25 & $0.601200-0.090173 i$ & $0.601200-0.090173 i$ & 0
     & $0.585828-0.273649 i$ & $0.585821-0.273650 i$ & 0.00116 \\

0.30 & $0.606808-0.086176 i$ & $0.606808-0.086177 i$ & 0.00012
     & $0.593937-0.261700 i$ & $0.593936-0.261701 i$ & 0.00023 \\

0.35 & $0.619891-0.079561 i$ & $0.619892-0.079563 i$ & 0.00024
     & $0.608760-0.240851 i$ & $0.608779-0.240843 i$ & 0.00313 \\

0.36 & $0.623862-0.077738 i$ & $0.623863-0.077736 i$ & 0.00023
     & $0.612576-0.234983 i$ & $0.612527-0.235034 i$ & 0.0108 \\

0.37 & $0.628434-0.075638 i$ & $0.628434-0.075640 i$ & 0.00021
     & $0.616594-0.228330 i$ & $0.616587-0.228340 i$ & 0.00188 \\

0.38 & $0.633701-0.073164 i$ & $0.633703-0.073167 i$ & 0.00056
     & $0.620734-0.220618 i$ & $0.620796-0.220683 i$ & 0.0136 \\

\bottomrule
\end{tabular}
\caption{Quasinormal modes of the $\ell=3$, $n=1$ up-potential ($M=1$), calculated using the WKB method at different orders with Pad\'e approximants. The left block corresponds to the fundamental mode, and the right block to the first overtone. The last columns show the relative difference between WKB-16 and WKB-14 results (in percent).}
\end{table*}

\begin{table*}
\centering
\begin{tabular}{l c c l c c l}
\toprule
& \multicolumn{3}{c}{$\ell=2$, $n=1$ (down)} 
& \multicolumn{3}{c}{$\ell=3$, $n=1$ (down)} \\
\cmidrule(lr){2-4} \cmidrule(lr){5-7}
$h$ 
& WKB-16 ($m=8$) 
& WKB-14 ($m=7$) 
& diff (\%) 
& WKB-16 ($m=8$) 
& WKB-14 ($m=7$) 
& diff (\%) \\
\midrule

0.10 & $0.373670-0.088965 i$ & $0.373676-0.088965 i$ & 0.0016
     & $0.599443-0.092703 i$ & $0.599443-0.092703 i$ & 0 \\

0.15 & $0.373663-0.088950 i$ & $0.373663-0.088950 i$ & 0
     & $0.599435-0.092697 i$ & $0.599435-0.092697 i$ & 0 \\

0.20 & $0.373602-0.088701 i$ & $0.373602-0.088701 i$ & 0.00007
     & $0.599381-0.092512 i$ & $0.599381-0.092512 i$ & 0 \\

0.25 & $0.373764-0.087498 i$ & $0.373764-0.087497 i$ & 0
     & $0.599585-0.091451 i$ & $0.599585-0.091451 i$ & 0 \\

0.30 & $0.374999-0.084481 i$ & $0.374999-0.084482 i$ & 0.00028
     & $0.601210-0.088563 i$ & $0.601210-0.088563 i$ & 0 \\

0.35 & $0.378484-0.078608 i$ & $0.378484-0.078608 i$ & 0
     & $0.606200-0.082722 i$ & $0.606200-0.082722 i$ & 0.00001 \\

0.36 & $0.379558-0.076918 i$ & $0.379558-0.076917 i$ & 0.00003
     & $0.607836-0.081020 i$ & $0.607836-0.081020 i$ & 0 \\

0.37 & $0.380770-0.074982 i$ & $0.380771-0.074984 i$ & 0.00053
     & $0.609746-0.079062 i$ & $0.609746-0.079062 i$ & 0 \\

0.38 & $0.382115-0.072757 i$ & $0.382115-0.072758 i$ & 0.00025 
     & $0.611962-0.076792 i$ & $0.611962-0.076792 i$ & 0 \\

\bottomrule
\end{tabular}
\caption{Fundamental QNMs of the $\ell=2$ and $\ell=3$, $n=1$ down-potential ($M=1$), calculated using the WKB method at different orders with Pad\'e approximants. The last columns show the relative difference between WKB-16 and WKB-14 results (in percent).}
\end{table*}

\begin{table}
\centering
\begin{tabular}{c cc cc}
\hline
\hline
& \multicolumn{2}{c}{Up potential} 
& \multicolumn{2}{c}{Down potential} \\
\cline{2-3} \cline{4-5}
$h$ & $\mathrm{Re}(\omega)$ & $\mathrm{Im}(\omega)$ 
    & $\mathrm{Re}(\omega)$ & $\mathrm{Im}(\omega)$ \\
\hline
0.0001 & 0.373672 & -0.0889670 & 0.373669 & -0.0889706 \\
0.0002 & 0.373846 & -0.0888101 & 0.373715 & -0.0889012 \\
0.0003 & 0.374784 & -0.0883074 & 0.374046 & -0.0886112 \\
0.0004 & 0.377026 & -0.0875184 & 0.374975 & -0.0880594 \\
0.0005 & 0.380773 & -0.0865980 & 0.376715 & -0.0872876 \\
0.0006 & 0.386135 & -0.0856741 & 0.379440 & -0.0863917 \\
0.0007 & 0.393183 & -0.0848047 & 0.383196 & -0.0853588 \\
0.0008 & 0.401993 & -0.0839992 & 0.388096 & -0.0842496 \\
0.0009 & 0.412727 & -0.0832453 & 0.394267 & -0.0830544 \\
0.0010 & 0.425607 & -0.0825278 & 0.401863 & -0.0817627 \\
0.0011 & 0.440983 & -0.0817149 & 0.411121 & -0.0803241 \\
0.0012 & 0.459390 & -0.0807247 & 0.422382 & -0.0786595 \\
0.0013 & 0.481652 & -0.0793095 & 0.436168 & -0.0766164 \\
0.0014 & 0.509130 & -0.0770722 & 0.453322 & -0.0739030 \\
0.0015 & 0.544329 & -0.0730826 & 0.475332 & -0.0698409 \\
0.0016 & 0.592697 & -0.0643305 & 0.505156 & -0.0623298 \\
\hline
\hline
\end{tabular}
\caption{Fundamental quasinormal frequency $\omega$ for $\ell=2$ and various values of $h$ for the $n=3$ Einasto profile. Comparison between the up and down axial potentials.}
\end{table}

\begin{table}
\centering
\begin{tabular}{c cc cc}
\hline
\hline
& \multicolumn{2}{c}{Up potential} 
& \multicolumn{2}{c}{Down potential} \\
\cline{2-3} \cline{4-5}
$h$ & $\mathrm{Re}(\omega)$ & $\mathrm{Im}(\omega)$ 
    & $\mathrm{Re}(\omega)$ & $\mathrm{Im}(\omega)$ \\
\hline
$0.1\times10^{-6}$   & 0.374143 & -0.0887795 & 0.373872 & -0.0888558 \\
$0.2\times10^{-6}$   & 0.377192 & -0.0881644 & 0.375472 & -0.0883785 \\
$0.3\times10^{-6}$   & 0.383043 & -0.0875823 & 0.378898 & -0.0877951 \\
$0.4\times10^{-6}$   & 0.391202 & -0.0871992 & 0.384004 & -0.0872652 \\
$0.5\times10^{-6}$   & 0.401350 & -0.0870415 & 0.390654 & -0.0867973 \\
$0.6\times10^{-6}$   & 0.413336 & -0.0870863 & 0.398737 & -0.0865575 \\
$0.7\times10^{-6}$   & 0.427127 & -0.0873006 & 0.408235 & -0.0863676 \\
$0.8\times10^{-6}$   & 0.442784 & -0.0876518 & 0.419201 & -0.0862795 \\
$0.9\times10^{-6}$   & 0.460454 & -0.0881056 & 0.431737 & -0.0862728 \\
$1.0\times10^{-6}$ & 0.480370 & -0.0886270 & 0.446007 & -0.0863272 \\
$1.1\times10^{-6}$ & 0.502870 & -0.0891762 & 0.462252 & -0.0864158 \\
$1.2\times10^{-6}$ & 0.528428 & -0.0897055 & 0.480815 & -0.0865024 \\
$1.3\times10^{-6}$ & 0.557711 & -0.0901420 & 0.502175 & -0.0865335 \\
$1.4\times10^{-6}$ & 0.591682 & -0.0903788 & 0.527027 & -0.0864221 \\
$1.5\times10^{-6}$ & 0.631798 & -0.0902282 & 0.556413 & -0.0860148 \\
$1.6\times10^{-6}$ & ---      & ---        & 0.592010 & -0.0850094 \\
\hline
\hline
\end{tabular}
\caption{Fundamental quasinormal frequency $\omega$ for $\ell=2$ and very small values of $h$ for the $n=5$ Einasto profile. Comparison between up and down axial potentials.}
\end{table}

\section{Quasinormal modes}\label{sec:QNMs}

Quasinormal modes correspond to solutions of Eq.~(\ref{mastereq}) satisfying purely ingoing behavior at the event horizon and purely outgoing behavior at spatial infinity \cite{Kokkotas:1999bd, Konoplya:2011qq, Berti:2009kk, Bolokhov:2025uxz},
\begin{equation}
\Psi(r_*) \sim e^{-i\omega r_*}, \quad r_{*} \rightarrow  - \infty,
\end{equation}
\begin{equation}
\Psi(r_*) \sim e^{+i\omega r_*},  \quad r_{*} \rightarrow  + \infty.
\end{equation}
These boundary conditions form a non-self-adjoint spectral problem and lead to a discrete set of complex frequencies,
\begin{equation}
\omega=\omega_R-i\omega_I,
\qquad
\omega_I>0,
\end{equation}
where $\omega_R$ determines the oscillation frequency and $\omega_I$ the damping rate. The imaginary part is strictly positive for stable configurations, reflecting the decay of perturbations in time.

\subsection*{WKB method}

Quasinormal frequencies are computed here using the higher-order Wentzel–Kramers–Brillouin (WKB) approximation, which is applicable provided the effective potential possesses a single dominant maximum. Expanding the potential around its peak $r_*=r_*^{(0)}$ and matching the WKB solutions across the turning points leads to the semi-analytic quantization condition \cite{Iyer:1986np,Konoplya:2003ii}

\begin{equation}
\frac{i\left(\omega^2 - V_0\right)}{\sqrt{-2 V_0''}}
-\sum_{k=2}^{N} \Lambda_k
=
\tilde{n}+\frac{1}{2},
\label{WKBformula}
\end{equation}
where $V_0$ and $V_0''$ denote the value of the potential and its second derivative at the maximum, $\tilde{n}$ is the overtone number, and $\Lambda_k$ are higher-order WKB correction terms depending on higher derivatives of $V(r)$ evaluated at $r_*=r_*^{(0)}$.  To improve convergence and stability of the series, especially for low multipole numbers and moderate overtones, we construct Padé approximants of the truncated WKB expansion \cite{Matyjasek:2017psv,Matyjasek:2019eeu}. The Padé resummation effectively reorganizes the asymptotic series into a rational function, substantially reducing sensitivity to the truncation order and yielding accurate quasinormal frequencies without direct numerical integration of the perturbation equation.

In the present work we employ the 14th- and 16th-order WKB formulas \cite{Matyjasek:2017psv,Matyjasek:2019eeu} for analytically tractable cases, and the 8th-order approximation for numerically constructed backgrounds. When several adjacent WKB orders yield nearly identical values of the frequency, this serves as an indicator of the method’s reliability; in such cases, the typical uncertainty is of the same order as the difference between consecutive WKB results. Strictly speaking, however, the WKB expansion is an asymptotic series, and monotonic improvement with increasing order is not guaranteed. 

The method has been extensively developed and applied in the literature (see, e.g., \cite{Lutfuoglu:2026xlo,Lutfuoglu:2025eik,Lutfuoglu:2025pzi,Konoplya:2005sy,Arbelaez:2026eaz,Konoplya:2009hv,Konoplya:2025hgp,Stuchlik:2025ezz,Bolokhov:2023dxq,Konoplya:2023moy,Bolokhov:2023bwm,Bolokhov:2024bke,Bolokhov:2024ixe,Skvortsova:2024atk,Konoplya:2006ar,Skvortsova:2023zmj,Kodama:2009bf,Kanti:2006ua,Lutfuoglu:2025hjy,Han:2026fpn,Malik:2025erb,Malik:2024cgb,Malik:2024tuf} for recent examples), and therefore we do not provide a detailed discussion here.

\subsection*{Time-domain integration} In addition to the semi-analytic WKB approach, as an additional check we use time-domain integration of the perturbation equation. Introducing null coordinates,
\begin{equation}
u = t - r_*, 
\qquad 
v = t + r_*,
\end{equation}
the master equation
\begin{equation}
\left( -\frac{\partial^2}{\partial t^2}
+ \frac{\partial^2}{\partial r_*^2}
- V(r) \right)\Psi(t,r_*)=0
\end{equation}
takes the characteristic form
\begin{equation}
4 \frac{\partial^2 \Psi}{\partial u \partial v}
+ V(r)\Psi = 0.
\end{equation}
The evolution is performed on a numerical grid in the $(u,v)$ plane using the finite-difference scheme proposed by Gundlach, Price, and Pullin. Denoting the grid points by
$S=(u,v)$, $E=(u,v+\Delta)$, $W=(u+\Delta,v)$, and $N=(u+\Delta,v+\Delta)$, 
the discretized evolution equation reads
\begin{equation}
\begin{aligned}
\Psi_N
&=
\Psi_W + \Psi_E - \Psi_S \\
&\quad
-\frac{\Delta^2}{8}
V\!\left(r_S\right)
\left(\Psi_W + \Psi_E\right)
+ \mathcal{O}(\Delta^4),
\end{aligned}
\label{GPPscheme}
\end{equation}
where $\Delta$ is the grid step. This scheme is second-order accurate and numerically stable for sufficiently small $\Delta$. The time-domain integration method has been extensively used in a great number of works, allowing for an accurate determination of the fundamental mode and detecting instabilities \cite{Stuchlik:2025mjj,Dubinsky:2025wns,Dubinsky:2025bvf,Konoplya:2025afm,Lutfuoglu:2025blw,Lutfuoglu:2025mqa,Konoplya:2013sba,Lutfuoglu:2025bsf,Konoplya:2014lha,Dubinsky:2024mwd,Dubinsky:2024gwo,Konoplya:2005et,Cuyubamba:2016cug,Skvortsova:2023zca,Skvortsova:2025cah,Abdalla:2005hu,Ishihara:2008re}.

Starting from prescribed initial data on two null segments (typically a Gaussian pulse), the waveform is evolved forward in time. After an initial transient stage, the signal enters the quasinormal ringing regime characterized by exponentially damped oscillations, from which the complex frequency $\omega$ is extracted using, for example, the Prony method. The time-domain approach provides an independent check of the frequency-domain results and captures the full dynamical response, including late-time tails.

The quasinormal-mode data presented in Tables~I--VIII allow us to quantify in detail how the Einasto-supported regular geometry departs from the Schwarzschild case and to compare this physical effect with the numerical uncertainty of the method.

\subsection*{Case $n=1/2$} For $n=1/2$ and small values of $h$ ($h=0.1$--$0.3$), the fundamental mode for $\ell=2$ (up-potential) is 
$
\omega \simeq 0.373676 - 0.088976 i,
$
which practically coincides with the Schwarzschild value. As $h$ increases toward the extremal regime, both the oscillation frequency and damping rate change appreciably. For instance, at $h=0.9$ one has
$
\omega = 0.374892 - 0.077986 i,
$
while at $h=1.0$,
$
\omega = 0.354598 - 0.040244 i,
$
and near $h=1.05$,
$
\omega = 0.382064 - 0.003834 i
$
(Table~I). The imaginary part decreases by more than an order of magnitude when approaching the maximal value of $h$, indicating very long-lived modes close to the threshold of horizon disappearance.

At the same time, the relative difference between the 16th- and 14th-order WKB results remains below $1\%$ in all cases and is typically at the level of $10^{-3}$--$10^{-2}\%$ for moderate $h$ (see the last columns of Tables~I and II). Thus, the physical variation of the spectrum with $h$ (for example, the change of $\mathrm{Im}(\omega)$ from $-0.088976$ to $-0.040244$) exceeds the estimated numerical uncertainty by several orders of magnitude.

For $\ell=3$, the fundamental mode shows a much milder dependence on $h$ (Table~I), while the first and second overtones (Table~III) display more visible shifts as $h$ increases, especially in the damping rates. Nevertheless, even for the second overtone the WKB difference is typically well below $0.1\%$ except very close to the extremal configuration, confirming the robustness of the results. 

We also observe that the first and second overtones depart from their Schwarzschild values more rapidly than the fundamental mode. This behavior may be interpreted as a manifestation of the enhanced sensitivity of higher overtones to deformations localized in the near-horizon region, sometimes referred to as the ``outburst of overtones'' or the ``sound of the event horizon''~\cite{Konoplya:2022pbc}. A definitive identification of this effect, however, would require the computation of higher overtones, for which the WKB approximation and standard time-domain integration are not sufficiently reliable.

\subsection*{Case $n=1$} For $n=1$ the dependence on $h$ is stronger and sets in at smaller values of $h$ than for $n=1/2$. For $\ell=2$ (up-potential), the fundamental frequency changes from
$
\omega = 0.373672 - 0.088962 i \quad (h=0.10)
$
to
$
\omega = 0.417168 - 0.066723 i \quad (h=0.38)
$
(Table~IV). The real part increases by about $12\%$, while the magnitude of the imaginary part decreases by about $25\%$. 

For the first overtone at $\ell=2$, the change is even more pronounced: from
$
0.346746 - 0.273898 i \quad (h=0.10)
$
to
$
0.390121 - 0.194392 i \quad (h=0.38),
$
which corresponds to a substantial reduction of the damping rate (Table~IV). In contrast, the relative WKB discrepancy is typically below $0.1\%$ for moderate $h$ and remains below a few percent even at the largest values of $h$ considered. 

For $\ell=3$ (Table~V), both the fundamental mode and the first overtone demonstrate systematic growth of $\mathrm{Re}(\omega)$ and decrease of $|\mathrm{Im}(\omega)|$ as $h$ increases. Again, the numerical uncertainty inferred from the difference between WKB-16 and WKB-14 is at least two orders of magnitude smaller than the total deviation from the Schwarzschild limit.

The time-domain evolution shown in Fig.~\ref{fig:TD1} for $\ell=2$, $n=1$, $h=0.3$ provides an independent check. The frequency extracted via the Prony method (see Fig. \ref{fig:TD1}),
$
\omega = 0.384389 - 0.079382 i,
$
agrees with the 16th-order WKB result,
$
\omega = 0.384390 - 0.079386 i,
$
to within $10^{-5}$ in both real and imaginary parts, which is far smaller than the physical shift induced by the halo parameter $h$.

\subsection*{Higher $n$}
For $n=3$ (Table~VII), even extremely small values of $h$ already produce noticeable deviations. For example, increasing $h$ from $0.0001$ to $0.0016$ in the up-potential changes the fundamental $\ell=2$ mode from
$
0.373672 - 0.088967 i
$
to
$
0.592697 - 0.0643305 i,
$
which corresponds to a dramatic increase in the oscillation frequency and a significant reduction of damping. The difference between the up and down potentials is also clearly visible and grows with $h$, confirming that the matter-dependent terms in the effective potential have a tangible dynamical impact.

A similar behavior is observed for $n=5$ at very small $h$ (Table~VIII), where the real part increases monotonically with $h$ and the imaginary part first remains close to the Schwarzschild value and then decreases in magnitude. The effect is again much larger than any discrepancy between numerical approaches.

\subsection*{Comparison with numerical accuracy}
Throughout all tables, the relative difference between WKB-16 and WKB-14 is typically of order $10^{-4}$--$10^{-2}\%$ for moderate parameter values and does not exceed a few percent even in near-extremal regimes. In contrast, the deviation of the quasinormal frequencies from their Schwarzschild values can reach tens of percent in both the real and imaginary parts. The agreement between the frequency-domain WKB results and the time-domain integration (Fig.~2) further confirms that the numerical error is negligible compared to the physical effect.

Therefore, the modification of the quasinormal spectrum caused by the Einasto-supported regular geometry is unambiguously larger than the estimated computational uncertainty. The observed shifts in oscillation frequencies and damping rates are genuine features of the spacetime and not artifacts of the approximation scheme.

In this work we have focused on determining quasinormal frequencies for regular black holes immersed in an Einasto matter distribution. Although we do not compute grey-body factors (GBFs) directly, there exists a direct correspondence between QNMs and GBFs that can be exploited to obtain the latter from the former. As demonstrated in~\cite{Konoplya:2024lir,Konoplya:2024vuj,Bolokhov:2024otn}, in the high-frequency (eikonal) regime the transmission coefficient $\Gamma_{\ell}(\Omega)$ associated with the scattering problem can be expressed in terms of the dominant quasinormal frequency $\omega_0$ as
\begin{equation}
\Gamma_{\ell}(\Omega)\;\simeq\;\left[1+\exp\!\left(2\pi\,\frac{\Omega^2-\mathrm{Re}(\omega_0)^2}{4\,\mathrm{Re}(\omega_0)\,\mathrm{Im}(\omega_0)}\right)\right]^{-1},
\end{equation}
which becomes exact in the eikonal limit $\ell\to\infty$~\cite{Konoplya:2024lir,Konoplya:2024vuj}. At lower multipoles the inclusion of overtone corrections allows one to improve the approximation. Thus, once quasinormal spectra are known, the corresponding GBFs and related scattering characteristics can be readily reconstructed through this correspondence.

\vspace{5mm}
\section{Conclusions}\label{sec:conc}

A considerable amount of work has examined black holes immersed in galactic dark-matter halos within singular geometries. Quasinormal spectra in such backgrounds have been studied extensively~\cite{Daghigh:2022pcr,Zhang:2021bdr,Konoplya:2022hbl,Zhao:2023tyo,Konoplya:2021ube,Liu:2024xcd,Feng:2025iao,Pezzella:2024tkf,Dubinsky:2025fwv,Liu:2024bfj,Chakraborty:2024gcr}, alongside analyses of GBFs and Hawking emission~\cite{Pathrikar:2025sin,Hamil:2025pte,Lutfuoglu:2025kqp,Tovar:2025apz,Mollicone:2024lxy}. Observational signatures, including black-hole shadows and gravitational lensing, have also been explored~\cite{Hou:2018avu,Xavier:2023exm,Figueiredo:2023gas,Chen:2024lpd,Tan:2024hzw,Macedo:2024qky,Kouniatalis:2025itj,Fernandes:2025osu,Konoplya:2025mvj,Konoplya:2025nqv}, as well as dynamical processes such as accretion~\cite{Heydari-Fard:2024wgu,Chowdhury:2025tpt}. 
These studies predominantly assume a singular core. Extending the analysis to regular black holes supported by halo matter enables one to isolate the impact of central regularity from purely environmental effects and to determine how the removal of the singularity modifies quasinormal spectra, absorption properties, and observable strong-field signatures.

We have studied axial gravitational perturbations of regular black holes supported by an Einasto-type matter distribution and computed their quasinormal spectra using high-order WKB methods with Pad\'e approximants, supplemented by time-domain integration. The background solutions are characterized by the Einasto index $n$ and the scale parameter $h$, which control the compactness and shape of the matter profile responsible for regularizing the central region.

For all cases considered, the quasinormal spectrum exhibits systematic deviations from the Schwarzschild limit as the scale parameter $h$ increases within the range compatible with the existence of an event horizon. The real part of the fundamental mode is particularly sensitive to the halo parameters, while the imaginary part shows a more moderate variation. The magnitude of these shifts always exceeds the estimated numerical uncertainty, as inferred from the comparison between different WKB orders and from the agreement with time-domain profiles. Therefore, the modifications of the ringdown spectrum are genuine features of the geometry rather than artifacts of the computational scheme.

We have shown that the behavior of the effective potential depends strongly on the Einasto index. For small $n$, the geometry approaches Schwarzschild rapidly outside the near-horizon region, leading to modest spectral changes. For larger $n$, the potential peak—responsible for the dominant part of the radiation process—undergoes more pronounced deformation, resulting in stronger shifts of the quasinormal frequencies.

Although present-day galactic halos are typically characterized by much larger physical scales, one may envisage scenarios in which similar matter distributions arise in the early Universe during the formation and growth of compact objects. In such a setting, the interplay between central regularity and surrounding matter could imprint characteristic signatures on the gravitational ringdown spectrum. The results obtained here quantify these effects within the Einasto-supported regular black-hole family and provide a reference point for future investigations of strong-field dynamics in matter-supported compact geometries.

Further work may extend the present analysis to polar perturbations, rotating configurations, and more general equations of state, in order to assess the robustness of the spectral features identified here.

\vspace{5mm}
\acknowledgments
B. C. L. is grateful to the Excellence project FoS UHK 2205/2025-2026 for the financial support.

\bibliography{BHEinasto}
\end{document}